\documentclass{ws-procs975x65}

\pdfoutput=1
\usepackage{color}
\usepackage{graphicx}
\usepackage{latexsym}
\usepackage{amssymb, amsmath, amsfonts}
\usepackage{indentfirst}
\usepackage{float,times}
\usepackage{bbm}
\usepackage{subfigure}
\usepackage{slashed}
\usepackage{ifthen}
\usepackage{mathrsfs}
\usepackage{setspace}

\newcommand{\drawsquare}[2]{\hbox{%
\rule{#2pt}{#1pt}\hskip-#2pt
\rule{#1pt}{#2pt}\hskip-#1pt
\rule[#1pt]{#1pt}{#2pt}}\rule[#1pt]{#2pt}{#2pt}\hskip-#2pt
\rule{#2pt}{#1pt}}
\newcommand{\Yfund}{\raisebox{-.5pt}{\drawsquare{6.5}{0.4}}}
\newcommand{\Yasymm}{\raisebox{-3.5pt}{\drawsquare{6.5}{0.4}}\hskip-6.9pt%
                     \raisebox{3pt}{\drawsquare{6.5}{0.4}}%
                    }
\newcommand{\Ysymm}{\Yfund\hskip-0.4pt%
                    \Yfund}

\def\drawbox#1#2{\hrule height#2pt
        \hbox{\vrule width#2pt height#1pt \kern#1pt
              \vrule width#2pt}
              \hrule height#2pt}

\def\Asym#1#2{\vcenter{\vbox{\drawbox{#1}{#2}
              \kern-#2pt 
              \drawbox{#1}{#2}}}}

\newcommand{\Ythreea}{\raisebox{-3.5pt}{\drawsquare{6.5}{0.4}}\hskip-6.9pt%
        \raisebox{3pt}{\drawsquare{6.5}{0.4}}\hskip-6.9pt
        \raisebox{9.5pt}{\drawsquare{6.5}{0.4}}}

\def\beq{\begin{equation}}
\def\eeq{\end{equation}}
\def\bal{\begin{align}}
\def\eal{\end{align}}

\newcommand{\Om}{{\cal O}_{\widetilde{\psi}{\psi}}}

\begin{document}
\begin{flushright}
\it CP$^3$-Origins:2010-7
\end{flushright}
\title{Phase Diagram of Strongly Interacting Theories}

\author{Francesco Sannino$^*$}

\address{CP$^3$-Origins, University of Southern Denmark,
Odense, 5230 M, Denmark\\
$^*$E-mail: sannino@cp3.sdu.dk, also sannino@cp3-origins.net \\
www.cp3-origins.dk}

\begin{abstract}
We summarize the phase diagrams of SU, SO and Sp gauge theories as function of  the number of flavors, colors, and matter representation as well as the ones of phenomenologically relevant chiral gauge theories such as the Bars-Yankielowicz and the generalized Georgi-Glashow models. We finally report on the intriguing possibility of the existence of gauge-duals for nonsupersymmetric gauge theories and the impact on their conformal window. 
\end{abstract}

\bodymatter

\section{Phases of Gauge Theories}
Models of dynamical breaking of the electroweak symmetry are theoretically appealing and constitute one of the best motivated natural extensions of the standard model (SM).  We have proposed several  models \cite{Sannino:2004qp,Dietrich:2005wk,Dietrich:2005jn,Gudnason:2006mk,Ryttov:2008xe,Frandsen:2009fs,Frandsen:2009mi,Antipin:2009ks} possessing interesting dynamics relevant for collider phenomenology \cite{Foadi:2007ue,Belyaev:2008yj,Antola:2009wq,Antipin:2010it} and cosmology \cite{Nussinov:1985xr,Barr:1990ca,Bagnasco:1993st,Gudnason:2006ug,Gudnason:2006yj,Kainulainen:2006wq,Kouvaris:2007iq,Kouvaris:2007ay,Khlopov:2007ic,Khlopov:2008ty,Kouvaris:2008hc,Belotsky:2008vh,Cline:2008hr,Nardi:2008ix,Foadi:2008qv,Jarvinen:2009wr,Frandsen:2009mi,Jarvinen:2009mh,Kainulainen:2009rb,Kainulainen:2010pk}. The structure of one of these models, known as Minimal Walking Technicolor, has led to the construction of a new supersymmetric extension of the SM featuring the maximal amount of supersymmetry in four dimension with a clear connection to string theory, i.e. Minimal Super Conformal Technicolor \cite{Antola:2010nt}. These models are also being investigated via first principle lattice simulations \cite{Catterall:2007yx,Catterall:2008qk,DelDebbio:2008zf,Hietanen:2008vc,Hietanen:2009az,Pica:2009hc,Catterall:2009sb,Lucini:2009an,Bursa:2009we,DeGrand:2009hu,DeGrand:2008kx,DeGrand:2009mt,Fodor:2008hm,Fodor:2009ar,Fodor:2009nh,Kogut:2010cz} \footnote{Earlier interesting models \cite{Appelquist:2002me,Appelquist:2003uu,Appelquist:2003hn} have contributed triggering the lattice investigations for the conformal
window with theories featuring fermions in the fundamental representation 
\cite{Appelquist:2009ty,Appelquist:2009ka,Fodor:2009wk,Fodor:2008hn,
Deuzeman:2009mh, Fodor:2009rb,Fodor:2009ff}}. An up-to-date review is Ref. \refcite{Sannino:2009za} while an excellent review updated till 2003 is Ref. \refcite{Hill:2002ap}.  These are also among the most challenging models to work with since they require deep knowledge of gauge dynamics in a regime where perturbation theory fails. In particular, it is of utmost importance to gain information on the nonperturbative dynamics of non-abelian four dimensional gauge theories.  The phase diagram of $SU(N)$ gauge theories as functions of number of flavors, colors and matter representation has been investigated in \cite{Sannino:2004qp,Dietrich:2006cm,Ryttov:2007sr,Ryttov:2007cx,Sannino:2008ha}.  The analytical tools which will be used here for such an exploration are: i) The conjectured {\it physical} all orders beta function for nonsupersymmetric gauge theories with fermionic matter in arbitrary representations of the gauge group \cite{Ryttov:2007cx}; ii) The truncated Schwinger-Dyson equation (SD) \cite{Appelquist:1988yc,Cohen:1988sq,Miransky:1996pd} (referred also as the ladder approximation in the literature);  The Appelquist-Cohen-Schmaltz (ACS) conjecture \cite{Appelquist:1999hr} which makes use of the counting of the thermal degrees of freedom at high and low temperature. These are the methods which we have used in our investigations. However several very interesting and competing analytic approaches \cite{Grunberg:2000ap,Gardi:1998ch,Grunberg:1996hu,Gies:2005as,Braun:2005uj, Poppitz:2009uq,Poppitz:2009tw,Antipin:2009wr,Antipin:2009dz,Jarvinen:2009fe, Braun:2009ns,Alanen:2009na} have been proposed in the literature. What is interesting is that despite the very different starting point  the various methods agree qualitatively on the main features of the various conformal windows presented here.  
\subsection{Physical all orders Beta Function - Conjecture}
\label{All-orders}
Recently we have conjectured an all orders beta function which allows for a bound of the conformal window \cite{Ryttov:2007cx} of $SU(N)$ gauge theories for any matter representation. The predictions of the conformal window coming from the above beta function are nontrivially supported by all the recent lattice results \cite{Catterall:2007yx,DelDebbio:2008wb,Catterall:2008qk,Appelquist:2007hu,
Shamir:2008pb,Deuzeman:2008sc,Lucini:2007sa}.

In \cite{Sannino:2009aw} we further assumed the form of the beta function to hold for $SO(N)$ and $Sp(2N)$ gauge groups and further extended in \cite{Sannino:2009za}  to chiral gauge theories. Consider a generic gauge group with $N_f(r_i)$ Dirac flavors belonging to the representation $r_i,\ i=1,\ldots,k$ of the gauge group. The conjectured beta function reads:
\begin{eqnarray}
\beta(g) &=&- \frac{g^3}{(4\pi)^2} \frac{\beta_0 - \frac{2}{3}\, \sum_{i=1}^k T(r_i)\,N_{f}(r_i) \,\gamma_i(g^2)}{1- \frac{g^2}{8\pi^2} C_2(G)\left( 1+ \frac{2\beta_0'}{\beta_0} \right)} \ ,
\end{eqnarray}
with
\begin{eqnarray}
\beta_0 =\frac{11}{3}C_2(G)- \frac{4}{3}\sum_{i=1}^k \,T(r_i)N_f(r_i) \qquad \text{and} \qquad \beta_0' = C_2(G) - \sum_{i=1}^k T(r_i)N_f(r_i)  \ .
\end{eqnarray}
The generators $T_r^a,\, a=1\ldots N^2-1$ of the gauge group in the
representation $r$ are normalized according to
$\text{Tr}\left[T_r^aT_r^b \right] = T(r) \delta^{ab}$ while the
quadratic Casimir $C_2(r)$ is given by $T_r^aT_r^a = C_2(r)I$. The
trace normalization factor $T(r)$ and the quadratic Casimir are
connected via $C_2(r) d(r) = T(r) d(G)$ where $d(r)$ is the
dimension of the representation $r$. The adjoint
representation is denoted by $G$.

The beta function is given in terms of the anomalous dimension of the fermion mass $\gamma=-{d\ln m}/{d\ln \mu}$ where $m$ is the renormalized mass, similar to the supersymmetric case \cite{Novikov:1983uc,Shifman:1986zi,Jones:1983ip}. 
The loss of asymptotic freedom is determined by the change of sign in the first coefficient $\beta_0$ of the beta function. This occurs when
\begin{eqnarray} \label{AF}
\sum_{i=1}^{k} \frac{4}{11} T(r_i) N_f(r_i) = C_2(G) \ , \qquad \qquad \text{Loss of AF.}
\end{eqnarray}
 At the zero of the beta function we have
\begin{eqnarray}
\sum_{i=1}^{k} \frac{2}{11}T(r_i)N_f(r_i)\left( 2+ \gamma_i \right) = C_2(G) \ ,
\end{eqnarray}
Hence, specifying the value of the anomalous dimensions at the IRFP yields the last constraint needed to construct the conformal window. Having reached the zero of the beta function the theory is conformal in the infrared. For a theory to be conformal the dimension of the non-trivial spinless operators must be larger than one in order not to contain negative norm states \cite{Mack:1975je,Flato:1983te,Dobrev:1985qv}.  Since the dimension of the chiral condensate is $3-\gamma_i$ we see that $\gamma_i = 2$, for all representations $r_i$, yields the maximum possible bound
\begin{eqnarray} 
\sum_{i=1}^{k} \frac{8}{11} T(r_i)N_f(r_i) = C_2(G) \ , \qquad \gamma_i = 2 \ .
\label{Bound}
\end{eqnarray}
In the case of a single representation this constraint yields 
\begin{equation}
N_f(r)^{\rm BF} \geq \frac{11}{8} \frac{C_2(G)}{T(r)} \ , \qquad \gamma  = 2 \ .
\end{equation}
The actual size of the conformal window can be smaller than the one determined by the bound above, Eq. (\ref{AF}) and (\ref{Bound}). It may happen, in fact, that chiral symmetry breaking is triggered for a value of the anomalous dimension less than two. If this occurs the conformal window shrinks. Within the ladder approximation \cite{Appelquist:1988yc,Cohen:1988sq} one finds that chiral symmetry breaking occurs when the anomalous dimension is close to one. Picking $\gamma_i =1$ we find:
\begin{eqnarray}
\sum_{i=1}^{k} \frac{6}{11} T(r_i)N_f(r_i) = C_2(G) \ , \qquad  \gamma = 1 \ ,. 
\end{eqnarray}
In the case of a single representation this constraint yields 
\begin{equation}
N_f(r)^{\rm BF} \geq \frac{11}{6} \frac{C_2(G)}{T(r)} \ , \qquad \gamma =1 \ .
\end{equation}
When considering two distinct representations the conformal window becomes a three dimensional volume, i.e. the conformal {\it house} \cite{Ryttov:2007sr}.  Of course, we recover the results by Banks and Zaks \cite{Banks:1981nn} valid in the perturbative regime of the conformal window.

We note that the presence of a physical IRFP requires the vanishing of the beta function for a certain value of the coupling. The opposite however is not necessarily
true; the vanishing of the beta function is not a sufficient condition to determine if the theory has a fixed point unless the beta function is {\it physical}. By {\it physical} we mean that the beta function allows to determine simultaneously other scheme-independent quantities at the fixed point such as the anomalous dimension of the mass of the fermions. This is exactly what our beta function does.  In fact, in the case of a single representation, one finds that at the zero of the beta function one has: 
\begin{eqnarray}
\gamma = \frac{11C_2(G)-4T(r)N_f}{2T(r)N_f} \ .\end{eqnarray}

 \subsection{Schwinger-Dyson in the Rainbow Approximation}
\label{ra}
{}For nonsupersymmetric theories another way to get quantitative estimates is to use the
{\it rainbow} approximation
to the Schwinger-Dyson equation
\cite{Maskawa:1974vs,Fukuda:1976zb}.  After a series of approximations (see \cite{Sannino:2009za} for a review)  one deduces for an $SU(N)$ gauge theory with $N_f$ Dirac fermions transforming according to the representation $r$ the critical number of flavors above which chiral symmetry maybe unbroken: 
\begin{eqnarray}
{N_f^{\rm SD}} &=& \frac{17C_2(G)+66C_2(r)}{10C_2(G)+30C_2(r)}
\frac{C_2(G)}{T(r)} \ . \label{nonsusy}
\end{eqnarray}
Comparing with the previous result obtained using the all orders beta function we see that it is the coefficient of $C_2(G)/T(r)$ which is different. We note that in \cite{Armoni:2009jn} it has been advocated a coefficient similar to the one of  the all-orders beta function.

\subsection{The $SU$, $SO$ and $Sp$ phase diagrams}
We consider here gauge theories with fermions in any representation of the $SU(N)$ gauge group \cite{Sannino:2004qp,Dietrich:2006cm,Ryttov:2007sr,Ryttov:2007cx,Ryttov:2009yw} using the various analytic methods described above. 

\begin{figure}[h]
\begin{center}\resizebox{10cm}{!}{\includegraphics{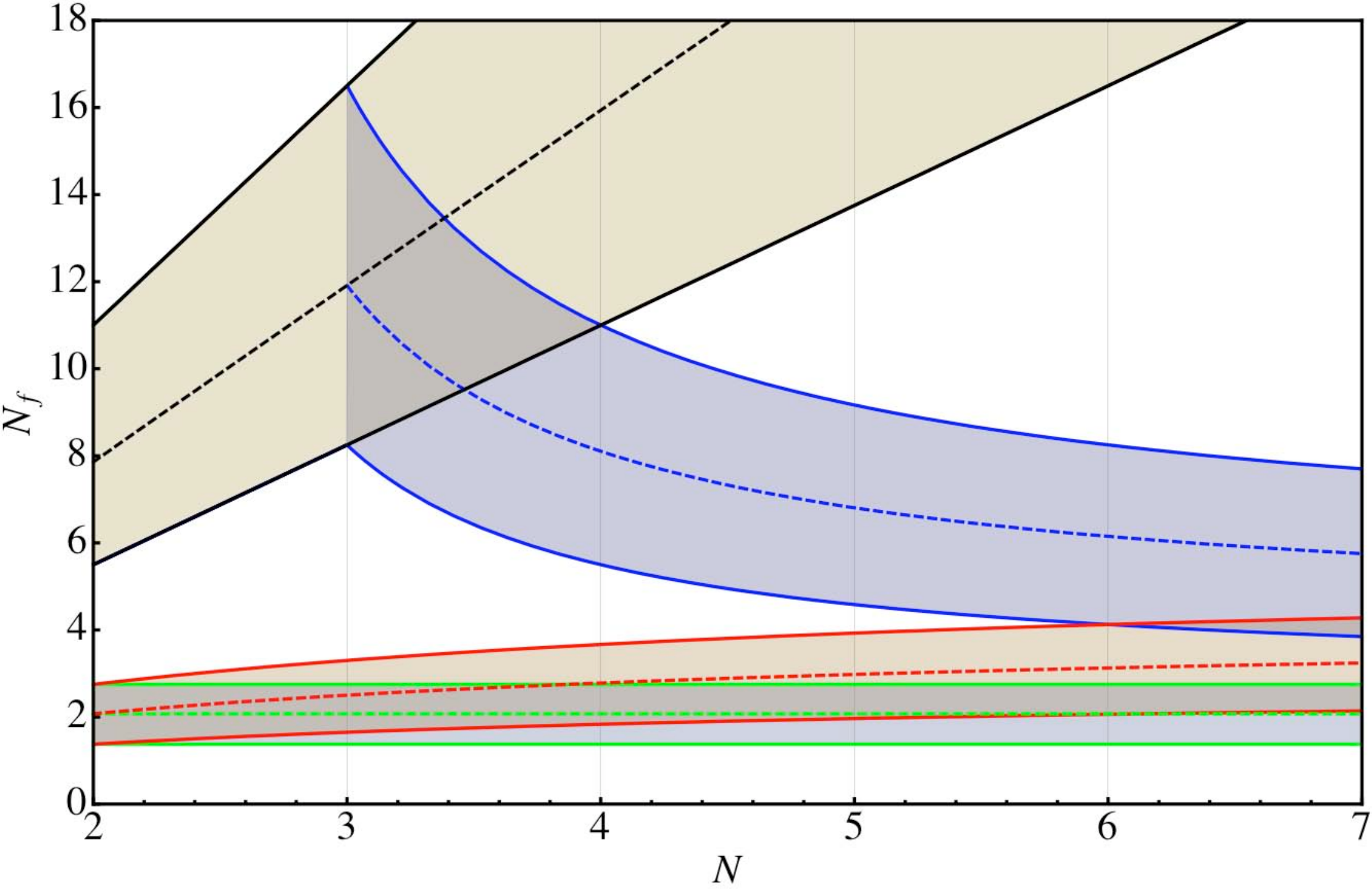}}
\caption{Phase diagram for nonsupersymmetric theories with fermions
in the: i) fundamental representation (black), ii) two-index
antisymmetric representation (blue), iii) two-index symmetric
representation (red), iv) adjoint representation (green) as a
function of the number of flavors and the number of colors. The
shaded areas depict the corresponding conformal windows. Above the
upper solid curve  the theories are no longer asymptotically free.
In between the upper and the lower solid curves the theories are
expected to develop an infrared fixed point according to the all orders
beta function. The area between the upper solid curve and
the dashed curve corresponds to the conformal window obtained in the
ladder approximation.} \label{PHComparison}\end{center}
\end{figure}

Here we plot in Fig.~\ref{PHComparison} the
conformal windows for various representations predicted with the physical all orders beta function and the SD approaches. 

The ladder result provides a size of the window, for every  fermion representation, smaller than the maximum bound found earlier. This is a consequence of the value of the anomalous dimension at the lower bound of the window. The unitarity constraint corresponds to $\gamma =2$ while the ladder result is closer to $\gamma \sim 1$. Indeed if we pick $\gamma =1$ our conformal window approaches the ladder result. Incidentally, a value of $\gamma$ larger than one, still allowed by unitarity, is a welcomed feature when using this window to construct walking technicolor theories. It may allow for the physical value of the mass of the top while avoiding a large violation of flavor changing neutral currents \cite{Luty:2004ye} which were investigated in  \cite{Evans:2005pu} in the case of the ladder approximation for minimal walking models.

\subsubsection{The $Sp(2N)$ phase diagram}
\label{sp}
$Sp(2N)$ is the subgroup of $SU(2N)$ which leaves the tensor
$J^{c_1 c_2} = ({\bf 1}_{N \times N} \otimes i \sigma_2)^{c_1 c_2}$
invariant. Irreducible tensors of $Sp(2N)$ must be traceless with respect to
$J^{c_1 c_2}$. 
Here we consider $Sp(2N)$ gauge theories with fermions transforming according to a given irreducible representation. Since $\pi^4\left[Sp(2N)\right] =Z_2$  there is a Witten topological anomaly \cite{Witten:1982fp} whenever the sum of the Dynkin indices of the various matter fields is odd. The adjoint of $Sp(2N)$ is the two-index symmetric tensor.

\begin{figure}[ht]
\centerline{
\includegraphics[height=6cm,width=11cm]{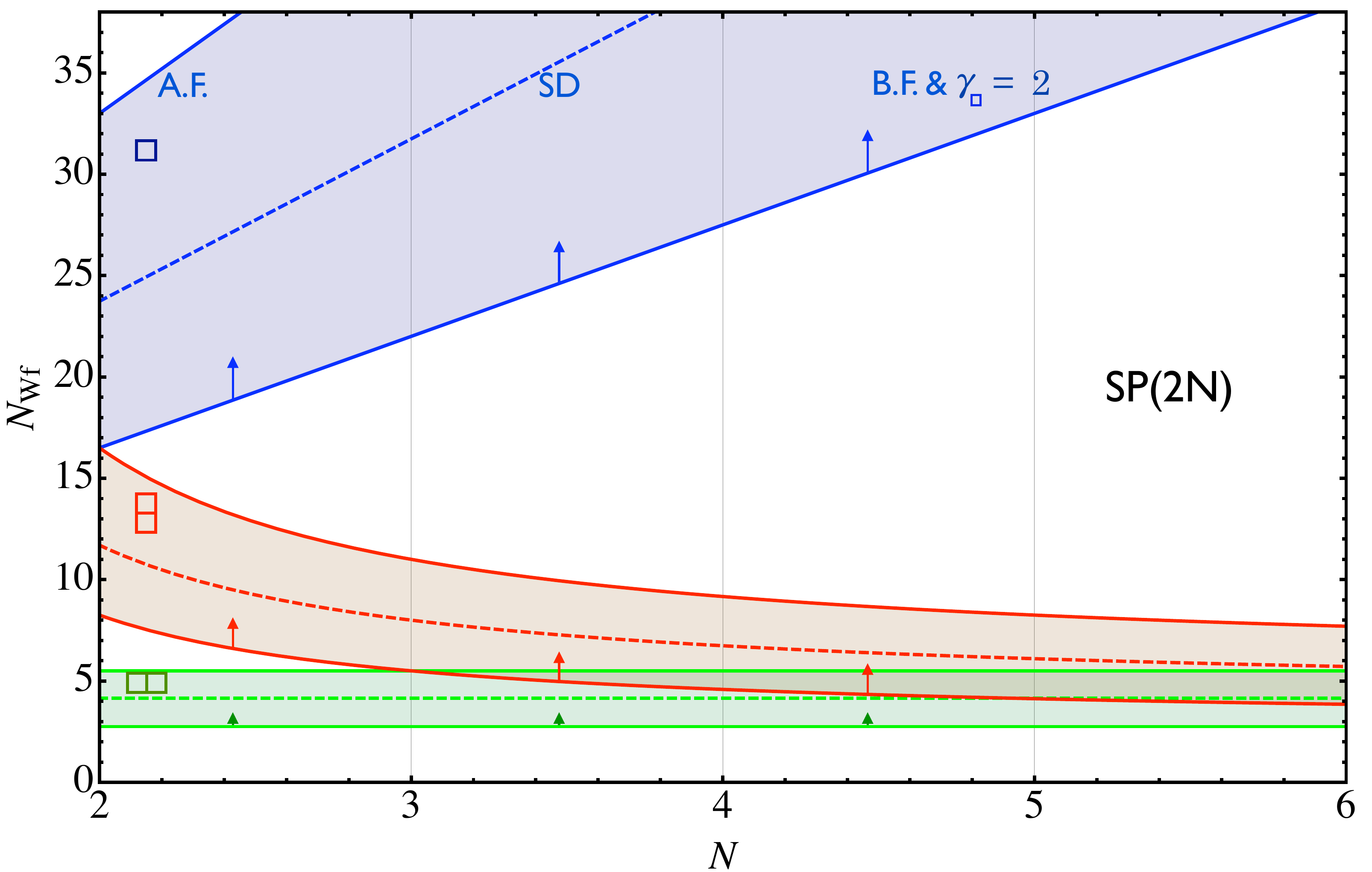}}
\caption
{Phase Diagram, from top to bottom, for $Sp(2N)$ Gauge Theories with $N_{Wf}=2N_f$  Weyl fermions in the vector representation (light blue),   $N_{Wf}=N_f$ in the  two-index antisymmetric representation (light red) and finally in the two-index symmetric (adjoint) (light green). The arrows indicate that the conformal windows can be smaller and the associated solid curves correspond to the all orders beta function prediction for the maximum extension of the conformal windows.} 
\label{Sp-PhaseDiagram}
\end{figure}

In Figure~\ref{Sp-PhaseDiagram} we summarize the relevant zero temperature and matter density phase diagram as function of the number of colors and Weyl flavors ($N_{Wf}$) for $Sp(2N)$ gauge theories. For the vector representation $N_{Wf} = 2N_f$ while for the two-index theories $N_{Wf} = N_f$. The shape of the various conformal windows are very similar to the ones for $SU(N)$ gauge theories \cite{Sannino:2004qp,Dietrich:2006cm,Ryttov:2007cx} with the difference that in this case the two-index symmetric representation is the adjoint representation and hence there is one less conformal window.

\subsubsection{The $SO(N)$ phase diagram}
\label{so}
We shall consider $SO(N)$ theories (for $N>5$) since they do not suffer of  a Witten
anomaly \cite{Witten:1982fp} and, besides, for $N<7$ can always be reduced to either an $SU$ or an $Sp$ theory.

\begin{figure}[t!]
\centerline{
\includegraphics[height=6cm,width=11cm]{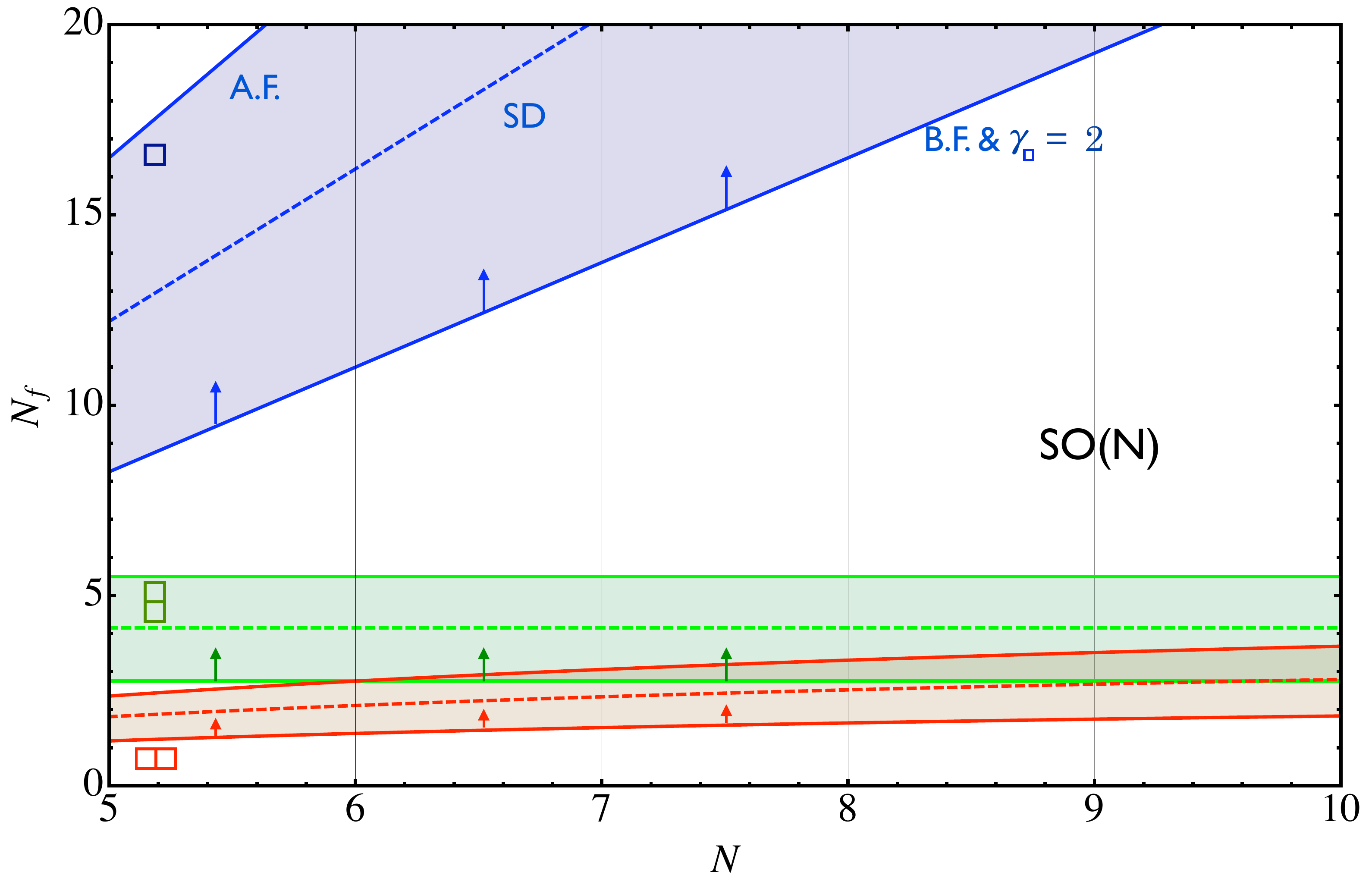}}
\caption
{Phase diagram of $SO(N)$ gauge theories with $N_f$  Weyl fermions in the vector representation, in the two-index antisymmetric (adjoint) and finally in the  two-index symmetric representation.  The arrows indicate that the conformal windows can be smaller and the associated solid curves correspond to the all orders beta function prediction for the maximum extension of the conformal windows. } 
\label{So-PhaseDiagram}
\end{figure}

In Figure~\ref{So-PhaseDiagram} we summarize the relevant zero temperature and matter density phase diagram as function of the number of colors and Weyl flavors ($N_{f}$) for $SO(N)$ gauge theories. The shape of the various conformal windows are very similar to the ones for $SU(N)$  and $Sp(2N)$ gauge  with the difference that in this case the two-index antisymmetric representation is the adjoint representation. We have analyzed only the theories with $N\geq 6$ since the remaining smaller $N$ theories can be deduced from $Sp$ and $SU$ using the fact that 
$SO(6)\sim SU(4)$, $SO(5)\sim Sp(4)$, 
$SO(4)\sim SU(2)\times SU(2)$, $SO(3)\sim SU(2)$, and $SO(2)\sim U(1)$.  

The phenomenological relevance of orthogonal gauge groups for models of dynamical electroweak symmetry breaking has been shown in \cite{Frandsen:2009mi}.

\subsection{Phases of Chiral Gauge Theories}

Chiral gauge theories, in which at least part of the matter field
content is in complex representations of the gauge group, play an
important role in efforts to extend the SM. These
include grand unified theories, dynamical breaking of symmetries,
and theories of quark and lepton substructure. Chiral theories received much attention in the 1980's~\cite{Ball:1988xg,Raby:1979my}.

Here we confront the  results obtained in Ref.~\cite{Appelquist:1999vs,Appelquist:2000qg} using the thermal degree of count freedom with the generalization of the all orders beta function useful to constrain chiral gauge theories appeared in \cite{Sannino:2009za}. The two important class of theories we are going to investigate are the Bars-Yankielowicz (BY) \cite{Bars:1981se} model involving fermions in the two-index symmetric tensor
representation, and the other is a generalized Georgi-Glashow (GGG)
model involving fermions in the two-index antisymmetric tensor
representation. In each case, in addition to fermions in complex
representations, a set of $p$ anti fundamental-fundamental pairs
are included and the allowed phases are considered as a function of
$p$. An independent relevant study of the phase diagrams of chiral gauge theories appeared in \cite{Poppitz:2009tw}. Here the authors also compare their results with the ones presented below.

\subsubsection{All-orders beta function for Chiral Gauge Theories}
A generic chiral gauge theory has always a set of matter fields for which one cannot provide a mass term, but it can also contain vector-like matter. We hence suggest the following minimal modification of the all orders beta function \cite{Ryttov:2007cx} for any nonsupersymmetric chiral gauge theory: 
\begin{equation}
\beta_{\chi} (g)= -\frac{g^3}{(4\pi)^2} \frac{\beta_0 - \frac{2}{3}\sum_{i=1}^{k}T(r_i)p(r_i) \gamma_i (g^2)} 
{1 - \frac{g^2}{8\pi^2}C_2(G)\left(1 + \frac{2\beta^{\prime}_{ \chi}}{\beta_0} \right)} \ ,
\end{equation}
where $p_i$ is the number of vector like pairs of fermions in the representation $r_i$ for which an anomalous dimension of the mass $\gamma_i$ can be defined.  $\beta_0$ is the standard one loop coefficient of the beta function while $\beta^{\prime}_{\chi}$ expression is readily obtained by imposing that when expanding  $\beta_{\chi}$ one recovers the two-loop coefficient correctly and its explicit expression is not relevant here.  According to the new beta function gauge theories without vector-like matter but featuring several copies of purely chiral matter will be conformal when the number of copies is such that the first coefficient of the beta function vanishes identically.  Using topological excitations an analysis of this case was performed in \cite{Poppitz:2009uq}. 
 
\subsubsection{ The Bars Yankielowicz (BY) Model}
\label{due}

This model is based on the single gauge group $SU(N\geq 3) $ and
includes fermions transforming as a symmetric tensor
representation, $S=\psi
_{L}^{\{ab\}}$, $a,b=1,\cdots ,N$; $\ N+4+p$ conjugate fundamental
representations: $\bar{F}_{a,i}=\psi _{a,iL}^{c}$, where $i=1,\cdots ,N+4+p$%
; and $p$ fundamental representations, $F^{a,i}=\psi _{L}^{a,i},\ i=1,\cdots
,p$. The $p=0$ theory is the basic chiral theory, free of gauge
anomalies by virtue of cancellation between the antisymmetric
tensor and the $N+4$ conjugate fundamentals. The additional $p$
pairs of fundamentals and conjugate fundamentals, in a real
representation of the gauge group, lead to no gauge anomalies.

The global symmetry group is
\begin{equation}
G_{f}=SU(N+4+p)\times SU(p)\times U_{1}(1)\times U_{2}(1)\ .
\label{gglobal3}
\end{equation}
Two $U(1)$'s are the linear combination of the original $U(1)$'s generated
by $S\rightarrow e^{i\theta _{S}}S$ , $\bar{F}\rightarrow e^{i\theta _{\bar{F%
}}}\bar{F}$ and $F\rightarrow e^{i\theta _{F}}F$ that are left invariant by
instantons, namely that for which $\sum_{j}N_{R_{j}}T(R_{j})Q_{R_{j}}=0$,
where $Q_{R_{j}}$ is the $U(1)$ charge of $R_{j}$ and $N_{R_{j}}$ denotes
the number of copies of $R_{j}$.

Thus the fermionic content of the theory is
\begin{table}[h]
\[ \begin{array}{c |  cc  c c c } \hline
{\rm Fields} &\left[ SU(N) \right] & SU(N+4+p) & SU(p) & U_1(1) &U_2(1) \\ \hline 
\hline 
 S &\Ysymm &1 &1 & N+4 &2p \\
  \bar{F} &\bar{\Yfund}  &\bar{\Yfund} & 1  & -(N+2) & - p \\
   {F} &\Yfund  &1 & \Yfund  & N+2 & - (N - p) \\
 \hline \end{array} 
\]
\caption{The Bars Yankielowicz (BY) Model}
\end{table}
where the first $SU(N)$ is the gauge group, indicated by the square brackets.

\begin{figure}[ht]
\centerline
{\includegraphics[height=4cm,width=16cm]{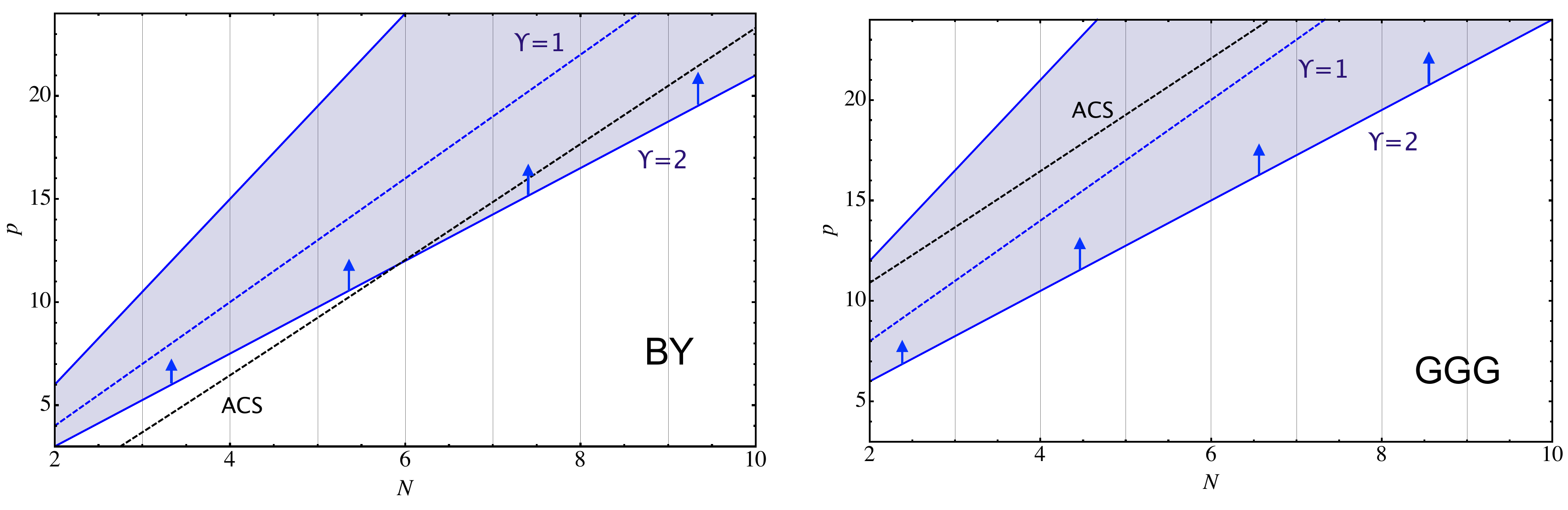}}
\caption
{ {\it Left panel}: Phase diagram of  the BY generalized model. The upper solid (blue) line corresponds to the loss of asymptotic freedom; the dashed (blue) curve corresponds to the chiral beta function prediction for the breaking/restoring of chiral symmetry. The dashed black line  corresponds to the ACS bound stating that the conformal region should start above this line. We have augmented the ACS method with the Appelquist-Duan-Sannino \cite{Appelquist:2000qg} extra requirement that the phase with the lowest number of massless degrees of freedom wins among all the possible phases in the infrared a chiral gauge theory can have. We hence used $f^{\rm brk + sym}_{IR}$ and $f_{UV}$ to determine this curve.  According to the all orders beta function (B.F.) the conformal window cannot extend below the solid (blue) line, as indicated by the arrows. This line corresponds to the anomalous dimension of the mass 
reaching the maximum value of $2$.  {\it Right panel}: The same plot for the GGG model.} 
\label{Chiral}
\end{figure}

From the numerator of the chiral beta function and the knowledge of the one-loop coefficient of the BY perturbative beta function the predicted conformal window is: 
\begin{equation}
 3\frac{(3N-2)}{2+\gamma^{\ast}}\leq p \leq \frac{3}{2}(3N-2) \ ,  
\end{equation}
with $\gamma^{\ast}$ the largest possible value of the anomalous dimension of the mass. The maximum value of the number of $p$ flavors is obtained by setting $\gamma^{\ast} = 2$: 
\begin{equation}
\frac{3}{4} (3N-2) \leq p \leq \frac{3}{2}(3N-2)  \ , \qquad \gamma^{\ast} = 2 \ ,
\end{equation}
while for $\gamma^{\ast} =1$ one gets: 
\begin{equation}
(3N-2) \leq p \leq \frac{3}{2} (3N-2)  \ , \qquad \gamma^{\ast} = 1 \ .
\end{equation}
The chiral beta function predictions for the conformal window are compared with the thermal degree of freedom investigation as shown in the left panel of Fig.~\ref{Chiral}.
In order to derive a prediction from the ACS method we augmented it with the Appelquist-Duan-Sannino \cite{Appelquist:2000qg} extra requirement that the phase with the lowest number of massless degrees of freedom wins among all the possible phases in the infrared a chiral gauge theory can have. The thermal critical number is: 
\begin{eqnarray}
p^{\rm Therm} = \frac{1}{4}\left[-16 + 3N + \sqrt{208  - 196N + 69 N^2} \right] \ .  
\end{eqnarray}

\subsubsection{ The Generalized Georgi-Glashow (GGG) Model}

This model is similar to the BY model just considered. It is an
$SU(N\geq 5)$ gauge theory, but with fermions in the
anti-symmetric, rather than symmetric, tensor representation. The
complete fermion content is $A=\psi _{L}^{[ab]},$ $a,b=1,\cdots
,N$; an additional $N-4+p$ fermions in the conjugate fundamental
representations: $\bar{F}_{a,i}=\psi
_{a,iL}^{c},$ $i=1,\cdots ,N-4+p$; and $p$ fermions in the fundamental
representations, $F^{a,i}=\psi _{L}^{a,i}$, $i=1,\cdots ,p$.

The global symmetry is
\begin{equation}
G_{f}=SU(N-4+p)\times SU(p)\times U_{1}(1)\times U_{2}(1) \ .
\end{equation}
where the two $U(1)$'s are anomaly free. With respect to this symmetry, the
fermion content is

\begin{table}[h]
\[ \begin{array}{c |  cc  c c c } \hline
{\rm Fields} &\left[ SU(N) \right] & SU(N -4+p) & SU(p) & U_1(1) &U_2(1) \\ \hline 
\hline &&&&&\\
 A &\Yasymm &1 &1 & N- 4 &2p \\
  \bar{F} &\bar{\Yfund}  &\bar{\Yfund} & 1  & -(N - 2) & - p \\
   {F} &\Yfund  &1 & \Yfund  & N - 2 & - (N - p) \\
 \hline \end{array} 
\]
\caption{The Generalized Georgi-Glashow (GGG) Model}
\end{table}

Following the analysis for the BY model the chiral beta function predictions for the conformal window are compared with the thermal degree of freedom investigation and the result is shown in the right panel of Fig.~\ref{Chiral}.

 \subsection{Conformal Chiral Dynamics}

\begin{figure}[h]
\begin{center}
\includegraphics[width=6cm]{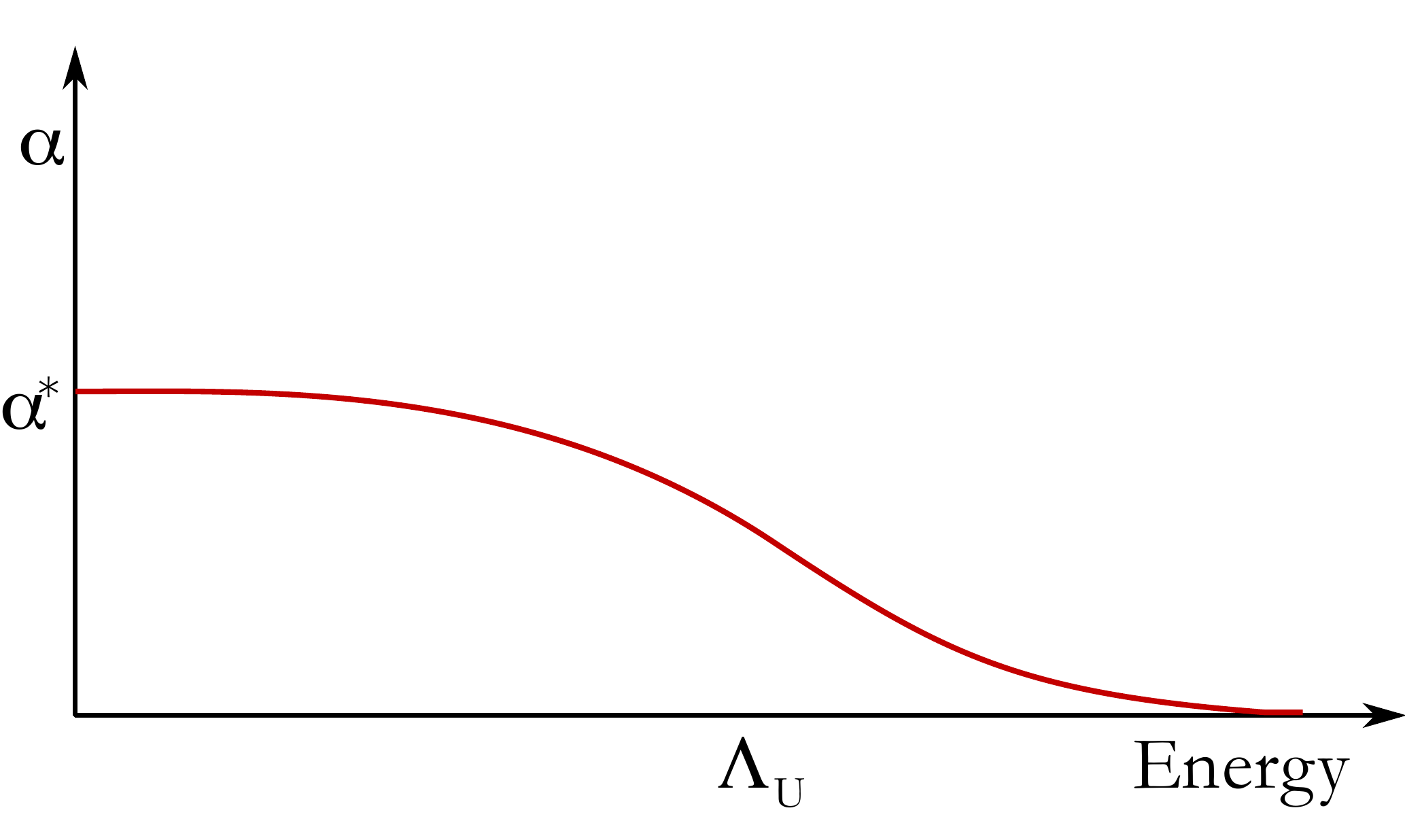}
\caption{ Running of the coupling constant in an asymptotically free gauge theory developing an infrared fixed point for a value $\alpha = \alpha^{\ast}$. }
\label{run1}
\end{center}
\end{figure}

Our starting point is a nonsupersymmetric non-abelian gauge theory with sufficient massless fermionic matter to develop a nontrivial IRFP. The cartoon of the running of the coupling constant is represented in Fig.~\ref{run1}. In the plot $\Lambda_U$ is the dynamical scale below which the IRFP is essentially reached. It can be defined as the scale for which $\alpha$ is $2/3$ of the fixed point value in a given renormalization scheme. 
If the theory possesses an IRFP the chiral condensate must vanish at large distances. Here we want to study the behavior of the condensate when a flavor singlet mass term is added to the underlying Lagrangian $ 
\Delta L = - m\,{\widetilde{\psi}}{\psi} + {\rm h.c.} $
with $m$ the fermion mass and $\psi^f_{c}$ as well as $\widetilde{\psi}_f^c$ left transforming two component spinors, $c$ and $f$ represent color and flavor indices.  The omitted color  and flavor indices, in the Lagrangian term, are contracted. 
We consider the case of fermionic matter in the fundamental representation of the $SU(N)$ gauge group. 
The effect of such a term is to break the conformal symmetry together with some of the global symmetries of the underlying gauge theory.  The composite operator $
{{\cal O}_{\widetilde{\psi}{\psi}}}^{f^{\prime}}_f  = \widetilde{\psi}^{f^{\prime}}{\psi}_f $
has mass dimension $\displaystyle{
d_{\widetilde{\psi}{\psi}} = 3 - \gamma}$ with 
$\gamma$  the anomalous dimension of the mass term. At the fixed point $\gamma$ is a positive number smaller than two \cite{Mack:1975je}. We assume $m \ll \Lambda_U$.  Dimensional analysis demands $
\Delta L  \rightarrow 
-m\, \Lambda_U^{\gamma} \, {\rm Tr}[\Om] + {\rm h.c.}~$.
The mass term is a relevant perturbation around the IRFP driving the theory away from the fixed point.  It will induce a nonzero vacuum expectation value for $\Om$ itself proportional to $\delta^{f^\prime}_f$. It is convenient to define ${\rm Tr}[\Om]  = N_f  \cal O $ with $\cal O$ a flavor singlet operator.  The relevant low energy Lagrangian term is then $  
-m\, \Lambda_U^{\gamma} \, N_f \cal O + {\rm h.c.} $. 
To determine the vacuum expectation value of $\cal{O}$ we follow \cite{Stephanov:2007ry,Sannino:2008nv}. 

The induced physical mass gap is a natural infrared cutoff. We, hence, identify  $\Lambda_{IR} $ with the physical value of the condensate. We find:
 \begin{eqnarray}
 \langle \widetilde{\psi}^f_c \psi^c_f \rangle  &\propto& -m \Lambda_U^2 \ ,  \qquad ~~~~~~0 <\gamma  < 1 \ , \label{BZm} \\
 \langle \widetilde{\psi}^f_c \psi^c_f \rangle  &\propto &   -m \Lambda_U^2  \log \frac{\Lambda^2_U}{|\langle {\cal O} \rangle|}\ , ~~~   \gamma \rightarrow  1    \ , \label{SDm} \\
 \langle \widetilde{\psi}^f_c\psi^c_f \rangle   &\propto &  -m^{\frac{3-\gamma} {1+\gamma}} 
 \Lambda_U^{\frac{4\gamma} {1+\gamma}}\ , ~~~1<\gamma  \leq 2 \ .
   \label{UBm}
 \end{eqnarray}
We used  $\langle \widetilde{\psi} \psi \rangle \sim \Lambda_U^{\gamma} \langle {\cal O} \rangle $ to relate the expectation value of ${\cal O}$ to the one of the fermion condensate. Via an allowed axial rotation $m$ is now real and positive. 
The effects of the Instantons on the conformal dynamics has been investigated in \cite{Sannino:2008pz}. Here it was shown that the effects of the instantons can be sizable only for a very small number of flavors given that, otherwise, the instanton induced operators are highly irrelevant.

\subsection {Gauge Duals and Conformal Window}

One of the most fascinating possibilities is that generic asymptotically free gauge theories have magnetic duals. In fact, in the late nineties, in a series of  ground breaking papers Seiberg  \cite{Seiberg:1994bz,Seiberg:1994pq} provided strong support for the existence of a consistent picture of such a duality within a supersymmetric framework. Arguably the existence of a possible dual of a generic nonsupersymmetric asymptotically free gauge theory able to reproduce its infrared dynamics must match the 't Hooft anomaly conditions. 

We have exhibited several solutions of these conditions for QCD in \cite{Sannino:2009qc} and for certain gauge theories with 
higher dimensional representations in  \cite{Sannino:2009me}. An earlier exploration already appeared in the literature \cite{Terning:1997xy}. The novelty with respect to these earlier results are: i) The request that the gauge singlet operators associated to the magnetic baryons should be interpreted as bound states of ordinary baryons \cite{Sannino:2009qc}; ii) The fact that the asymptotically free condition for the dual theory matches the lower bound on the conformal window obtained using the all orders beta function  \cite{Ryttov:2007cx}. These extra constraints help restricting further the number of possible gauge duals without diminishing the exactness of the associate solutions with respect to the 't Hooft anomaly conditions.

We will briefly summarize here the novel solutions to the 't Hooft anomaly conditions for QCD. The resulting {\it magnetic} dual allows to predict the critical number of flavors above which the asymptotically free theory, in the electric variables, enters the conformal regime as predicted using the all orders conjectured beta function \cite{Ryttov:2007cx}.

\subsubsection{QCD Duals} 
The underlying gauge group is $SU(3)$ while the
quantum flavor group is
\begin{equation}
SU_L(N_f) \times SU_R(N_f) \times U_V(1) \ ,
\end{equation}
and the classical $U_A(1)$ symmetry is destroyed at the quantum
level by the Adler-Bell-Jackiw anomaly. We indicate with
$Q_{\alpha;c}^i$ the two component left spinor where $\alpha=1,2$
is the spin index, $c=1,...,3$ is the color index while
$i=1,...,N_f$ represents the flavor. $\widetilde{Q}^{\alpha ;c}_i$
is the two component conjugated right spinor.

The  global anomalies are associated to the triangle diagrams featuring at the vertices three $SU(N_f)$ generators (either all right or all left), or two 
$SU(N_f)$ generators (all right or all left) and one $U_V(1)$ charge. We indicate these anomalies for short with:
\begin{equation}
SU_{L/R}(N_f)^3 \ ,  \qquad  SU_{L/R}(N_f)^2\,\, U_V(1) \ .
\end{equation}
For a vector like theory there are no further global anomalies. The
cubic anomaly factor, for fermions in fundamental representations,
is $1$ for $Q$ and $-1$ for $\tilde{Q}$ while the quadratic anomaly
factor is $1$ for both leading to
\begin{equation}
SU_{L/R}(N_f)^3 \propto \pm 3 \ , \quad SU_{L/R}(N_f)^2 U_V(1)
\propto \pm 3 \ .
\end{equation}

\begin{figure}[h!]
\centerline{\includegraphics[width=8cm]{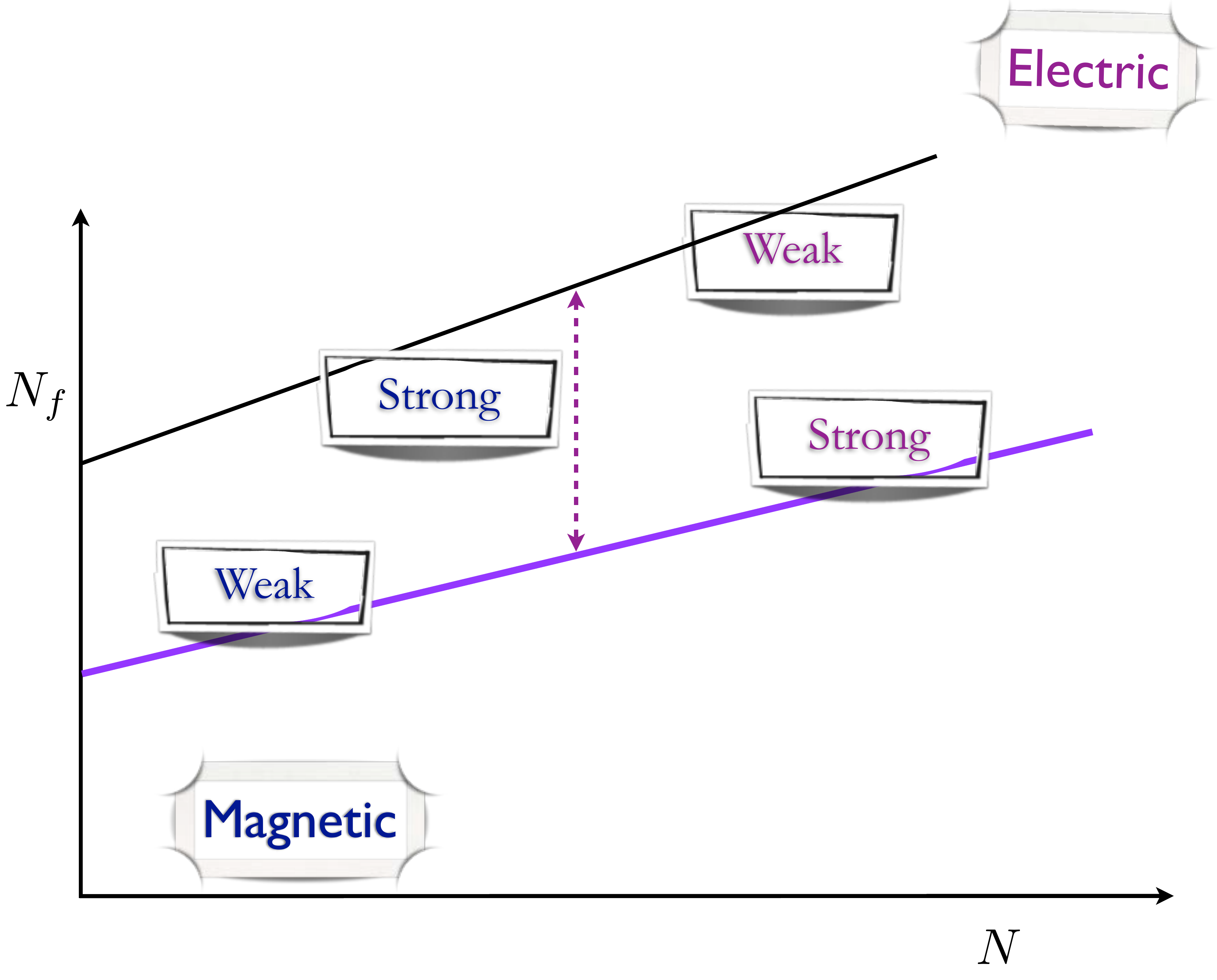}}
\caption{Schematic representation of the phase diagram as function of number of flavors and colors. For a given number of colors by increasing the number flavors within the conformal window we move from the lowest line (violet) to the upper (black) one. The upper black line corresponds to the one where one looses asymptotic freedom in the electric variables and the lower line where chiral symmetry breaks and long distance conformality is lost. In the {\it magnetic} variables the situation is reverted and the perturbative line, i.e. the one where one looses asymptotic freedom in the magnetic variables, correspond to the one where chiral symmetry breaks in the electric ones. }
\label{Duality}
\end{figure}

If a magnetic dual of QCD does exist one expects it to be weakly coupled near the critical number of flavors below which one breaks  large distance conformality in the electric variables.  This idea is depicted in Fig~\ref{Duality}. 

Determining a possible unique dual theory for QCD is, however, not simple given the few mathematical constraints at our disposal. The saturation of the global anomalies is an important tool but is not able to select out a unique solution. We shall see, however, that one of the solutions, when interpreted as the QCD dual, leads to a prediction of a critical number of flavors corresponding exactly to the one obtained via the conjectured all orders beta function.

 We seek solutions of the anomaly matching conditions for a gauge theory $SU(X)$ with global symmetry group $SU_L(N_f)\times SU_R(N_f) \times U_V(1)$  featuring 
{\it magnetic} quarks ${q}$ and $\widetilde{q}$ together with $SU(X)$ gauge singlet states identifiable as baryons built out of the {\it electric} quarks $Q$. Since mesons do not affect directly global anomaly matching conditions we could add them to the spectrum of the dual theory.  We study the case in which $X$ is a linear combination of number of flavors and colors of the type $\alpha N_f + 3 \beta$ with $\alpha$ and $\beta$ integer numbers. 

We add to the {\it magnetic} quarks gauge singlet Weyl fermions which can be identified with the baryons of QCD but massless. 

Having defined the possible massless matter content of the gauge theory dual to QCD one computes the $SU_{L}(N_f)^3$ and $SU_{L}(N_f)^2\,\, U_V(1)$ global anomalies in terms of the new fields.   
We have found several solutions to the anomaly matching conditions presented above. Some were found previously in \cite{Terning:1997xy}. Here we display a new solution in which the gauge group is $SU(2N_f - 5N)$  with the number of colors $N$  equal to $3$. It is, however, convenient to keep the dependence on $N$ explicit. 
 \begin{table}[bh]
\[ \begin{array}{|c| c|c c c|c|} \hline
{\rm Fields} &\left[ SU(2N_f - 5N) \right] & SU_L(N_f) &SU_R(N_f) & U_V(1)& \# ~{\rm  of~copies} \\ \hline 
\hline 
 q &\Yfund &{\Yfund }&1& \frac{N(2 N_f - 5)}{2 N_f - 5N} &~~~1 \\
\widetilde{q} & \overline{\Yfund}&1 &  \overline{\Yfund}& -\frac{N(2 N_f - 5)}{2 N_f - 5N}&~~~1   \\
A &1&\Ythreea &1&~~~3& ~~~2 \\
B_A &1&\Yasymm &\Yfund &~~~3& -2\\
{D}_A &1&{\Yfund} &{\Yasymm } &~~~3& ~~~2 \\
\widetilde{A} &1&1&\overline{\Ythreea} &-3&~~~2\\
 \hline \end{array} 
\]
\caption{Massless spectrum of {\it magnetic} quarks and baryons and their  transformation properties under the global symmetry group. The last column represents the multiplicity of each state and each state is a  Weyl fermion.}
\label{dual}
\end{table}
$X$ must assume a value strictly larger than one otherwise it is an abelian gauge theory. This provides the first nontrivial bound on the number of flavors: 
 \begin{equation}
 N_f > \frac{5N + 1}{2}  \ , \end{equation} 
which for $N=3$ requires $N_f> 8 $. 
Asymptotic freedom of the newly found theory is dictated by the coefficient of the one-loop beta function :
 \begin{equation}
 \beta_0 = \frac{11}{3} (2N_f - 5N)  - \frac{2}{3}N_f \ . 
 \end{equation}
 To this order in perturbation theory the gauge singlet states do not affect the {magnetic} quark sector and we can hence determine  the number of flavors obtained by requiring the dual theory to be asymptotic free. i.e.: 
\begin{equation}
N_f \geq \frac{11}{4}N \qquad\qquad\qquad {\rm Dual~Asymptotic~Freedom}\ . 
\end{equation}
Quite remarkably this value {\it coincides} with the one predicted by means of the all orders conjectured beta function for the lowest bound of the conformal window, in the {\it electric} variables, when taking the anomalous dimension of the mass to be $\gamma =2 $. We recall that for any number of colors $N$ the all orders beta function requires the critical number of flavors to be larger than: 
\begin{equation}
N_f^{BF}|_{\gamma = 2} = \frac{11}{4} N \ . 
\end{equation}
{}For N=3 the two expressions yield $8.25$ \footnote{Actually given that $X$ must be at least $2$ we must have  $N_f \geq 8.5$ rather than $8.25$}. We consider this a nontrivial and  interesting result lending further support to the all orders beta function conjecture and simultaneously suggesting that this theory might, indeed, be the QCD magnetic dual.
 The actual size of the conformal window matching this possible dual corresponds to setting $\gamma =2$. {}We note that although for $N_f = 9$ and $N=3$ the magnetic gauge group is $SU(3)$ the theory is not trivially QCD given that it features new massless fermions and their interactions with massless mesonic type fields.

Recent suggestions to analyze the conformal window of nonsupersymmetric gauge theories based on different model assumptions \cite{Poppitz:2009uq} are in qualitative agreement with the precise results of the all orders beta function conjecture. It is worth noting that the combination $2N_f - 5N$ appears in the computation of the mass gap for gauge fluctuations presented in \cite{Poppitz:2009uq,Poppitz:2008hr}. It would be interesting to explore a possible link between these different approaches in the future.

We have also find solutions for which the lower bound of the conformal window is saturated for $\gamma =1$. The predictions from the gauge duals are, however, entirely and surprisingly consistent with the maximum extension of the conformal window obtained using the all orders beta function \cite{Ryttov:2007cx}. Our main conclusion is that the  't Hooft anomaly conditions alone do not exclude the possibility that the maximum extension of the QCD conformal window is the one obtained for a large anomalous dimension of the quark mass.  

By computing the same gauge singlet correlators in QCD and its suggested dual, one can directly validate or confute this proposal via lattice simulations.

\subsection{Conclusions}

We investigated the conformal windows of chiral and non-chiral nonsupersymmetric gauge theories with fermions in any representation of the underlying gauge group using four independent analytic methods. {}For vector-like gauge theories one observes a universal value, i.e. independent of the representation, of the ratio of the area of the maximum extension of the conformal window, predicted using the all orders beta function, to the asymptotically free one, as defined in \cite{Ryttov:2007sr}. It is easy to check from the results presented that this ratio is not only independent on the representation but also on the particular gauge group chosen. 

The four  methods we used to unveil the conformal windows are the all orders beta function (BF), the SD truncated equation, the thermal degrees of freedom method and possible gauge - duality. They have vastly different starting points and there was no, a priori, reason to agree with each other, even at the qualitative level.

Several questions remain open such as what happens on the right hand side of the infrared fixed point as we increase further the coupling. Does a generic strongly coupled theory develop a new UV fixed point as we increase the coupling beyond the first IR value \cite{Kaplan:2009kr}? If this were the case our beta function would still be a valid description of the running of the coupling of the constant in the region between the trivial UV fixed point and the neighborhood of the first IR fixed point. One might also consider extending our beta function to take into account of this possibility as done in \cite{Antipin:2009wr}. It is also possible that no non-trivial UV fixed point forms at higher values of the coupling constant for any value of the number of flavors within the conformal window. Gauge-duals seem to be in agreement with the simplest form of the beta function. The extension of the all orders beta function to take into account fermion masses has appeared in \cite{Dietrich:2009ns}.

Our analysis  substantially increases the number of asymptotically free gauge theories which can be used to construct SM extensions making use of (near) conformal dynamics. Current Lattice simulations can test our predictions and lend further support or even disprove  the emergence of a universal picture possibly relating the phase diagrams of gauge theories of fundamental interactions.

\subsubsection*{Acknowledgments}
I would like to thank the organizers of SCGT09 for providing a very interesting scientific meeting. It is a pleasure to thank M. Antola, T. Appelquist, S. Catterall, L. Del Debbio, S. Di Chiara, D.D. Dietrich, J. Giedt, R. Foadi, M.T. Frandsen, M. J\"arvinen, C. Pica, T.A. Ryttov, M. Shifman, J. Schechter. R. Shrock, and K. Tuominen who have contributed getting me interested in this interesting subject, substantially helped developing the ideas and the results reported here and finally for discussions and comments. 



\end{document}